\documentclass[useAMS,usenatbib]{mn2e} \usepackage{epsfig}
\usepackage{amsmath}
\usepackage{natbib}

\author[Theis, Morgan, and Fortenberry]
{Riley A. Theis, W. James Morgan, and Ryan C.
Fortenberry\thanks{Email:rfortenberry@georgiasouthern.edu}
\\
Georgia Southern University, Department of Chemistry, Statesboro, GA 30460
U.S.A.}

\title[NeH$_2$$^+$ and ArH$_2$$^+$]{ArH$_2$$^+$ and NeH$_2$$^+$ as
Global Minima in the Ar$^+$/Ne$^+$ + H$_2$ Reactions: Energetic, Spectroscopic,
and Structural Data}

\begin{document}

\date{Submitted: \today}

\maketitle

\begin{abstract}

In light of the recent discovery of ArH$^+$ in the Crab nebula, it is shown
through high-level quantum chemical comptuations that the global minima on the
Ar$^+$/Ne$^+$ + H$_2$ potential energy surfaces are ArH$_2$$^+$ and
NeH$_2$$^+$.  Hence, ArH$_2$$^+$ may be a necessary intermediate in the Ar$^+ +
\mathrm{H}_2 \rightarrow \mathrm{ArH}^+ + \mathrm{H}$ formation reaction
proposed in the same work where ArH$^+$ is first reported in the Crab nebula.
ArH$_2$$^+$ is also probably an intermediate in the alternative $\mathrm{Ar} +
\mathrm{H}_2^+ \rightarrow \mathrm{ArH}^+ + \mathrm{H}$ reaction.
Additionally, it is shown that Ne$^+ + \mathrm{H}_2 \rightarrow
\mathrm{NeH}_2^+$ will subsequently most likely yield $\mathrm{Ne} +
\mathrm{H}_2^+$ and not NeH$^+ + \mathrm{H}$ offering a possible rationale as
to the absence of NeH$^+$ in spectra obtained from the interstellar medium
(ISM).  Following from this, spectroscopic data (both rotational and
vibrational) are provided for NeH$_2$$^+$ and ArH$_2$$^+$ through the use of
highly-accurate quantum chemical quartic and cubic force fields.  All possible
isotopologues are also included for $^{20}$Ne, $^{22}$Ne, $^{36}$Ar, $^{38}$Ar,
$^{40}$Ar, $^1$H, and D.  The dipole moments for these systems are quite large
at 5.61 D for NeH$_2$$^+$ and 4.37 D for ArH$_2$$^+$.  The spectroscopic
constants provided will aid in the potential detection of these open-shell
noble gas dihydride cations in the ISM.

\begin{keywords}
astrochemistry -- molecular data -- molecular processes -- ISM: molecules --
radio lines: ISM -- infrared: ISM
\end{keywords}

\end{abstract}

\section{Introduction}

The recent detection of the noble gas compound $^{36}$ArH$^+$ in the Crab
nebula \citep{Barlow13} and, more recently in various other regions
\citep{Schilke14}, has tremendously opened up the chemistry of the interstellar
medium (ISM), and has led to questions about the provenance of this molecule
\citep{Roueff14} and the possible existence of other noble gas compounds to be
found in various regions of the ISM.  ArH$^+$ has been proposed to form via the
Ar$^+ + \mathrm{H}_2$ reaction in the ISM \citep{Barlow13} and has been
explored both observationally and theoretically \citep{Schilke14, Roueff14}.
The noble gases (He, Ne, Ar, Kr, Xe) have been called ``noble'' due to their
aversion to chemical reactivity brought about by the complete filling of their
valence electron shell.  Hence, electrons are not easily shared by these atoms
and bonds are not typically made.  However, noble gas atoms can be polarized by
various substituent groups, and cationic and van der Waals noble gas compounds
have long been studied \citep{Grandinetti11}.  What is intriguing and
ground-breaking about the discovery of ArH$^+$ in the Crab nebula is that a
noble gas compound has finally been shown to exist somewhere in nature.

The ISM is a natural place to find noble gas compounds where the pressures and
temperatures of a given region may be just right for the synthesis of
terrestrially unstable molecules.  Radicals and even anions have been commonly
detected in the ISM by comparison, largely, to spectroscopic data made
available in databases like the Cologne Database for Molecular Spectroscopy
(CDMS) \citep{Cologne05}. However, the chief problem with the detection of noble
gas compounds in the ISM is a lack of adequate spectroscopic data since these
molecules are difficult to study in the laboratory combined with the ensuing
fact that little need has previously been demonstrated for such data.  A
follow-up study \citep{Cueto14} to this initial discovery has produced
highly-accurate rovibrational reference data for $^{36}$ArH$^+$, as well as the
other abundant Ar isotope $^{38}$ArH$^+$, in order to augment that from the
CDMS which was used in the detection of $^{36}$ArH$^+$ in the Crab nebula.
Hence, if more noble gas compounds are to be detected, reliable spectroscopic
data must be generated.

Searches for the related HeH$^+$ and NeH$^+$ are almost certainly underway if
not already completed \citep{Schilke14}.  It is unlikely that noble gas
compounds containing krypton and xenon will be observed since these atoms are
both larger than iron and exist in substantially smaller abundances.  However,
the large dipole moments exhibited by noble gas hydride cations \citep{Cueto14}
brought about from the drastic difference between the centre-of-mass (very
close to the noble gas atomic centre) and the centre-of-charge (roughly midway
between the two atoms) make them very rotationally bright aiding in their
potential detection as a class.  Even so, other compounds containing He, Ne,
and Ar are more likely to be observed in the ISM than those containing the
heavier noble gas atoms.

Various noble gas species, both cationic and neutral, have been shown to be
stable through laboratory experiments and quantum chemical computations.  For
example, high-resolution infrared data of the Ng-H$_2$ (Ng = Ne, Ar, Kr)
structures have been provided \citep{McKellar96, McKellar05, McKellar09}, and
computational studies as to their energy profiles, as well as that of HeH$_2$,
have shown that these are minima on their respective potential energy surfaces
(PESs) \citep{Barletta09}. Additionally, complexes of one to three noble gas
atoms bonded to H$_3$$^+$, a molecule of substantial importance to
astrochemistry \citep{Pavanello12}, have exhibited marked
stabilities \citep{Pauzat09}.  Other noble gas molecules are known \citep{Kim99,
Borocci11, Grandinetti11}, but the next-simplest class of noble gas compounds
are simply the Ng-H$_2$$^+$ species.

There exists very little data on this class of compounds which is surprising
due to their relative simplicity.  \citet{Bartl13} recently observed
HeH$_2$$^+$, in addition to ``all conceivable combinations of $n$ and $x$'' for
He$_n$H$_x$$^+$, in mass spectrometry experiments, indicating that these
Ng-H$_2$$^+$ structures are likely stable minima, as well.  \cite{Gamallo13}
studied the reaction of Ne + H$_2$$^+$ and found that the global minimum on the
PES is actually NeH$_2$$^+$ following work on this same surface from
\cite{Mayneris08}.  Several studies have explored the interaction of argon
atoms and atomic cations with hydrogen molecules and molecular cations
\citep{Bedford90, Tosi93, Song03} where most of these build on the seminal work
by \cite{Adams70} where noble gas dimers were reacted with hydrogen molecules.
ArH$_2$$^+$ is believed to exist on the PES due to the relative energetics of
the total surface, and it seems probable, therefore, that these Ng-H$_2$$^+$
molecules are experimentally observable.  However, spectroscopic data has not
been provided for the Ng-H$_2$$^+$ molecules in order to confirm their presence
in these experiments.

In this present study, the formation/dissociation pathways of two Ng-H$_2$$^+$
cations, NeH$_2$$^+$ and ArH$_2$$^+$, are provided.  Additionally, the
rovibrational spectroscopic data of these two noble gas dihydride cations are
produced in order to assist in their potential discovery in the ISM.  Building
on previous quantum chemical studies that have generated rotational constants to
better than $0.1\%$ of the corresponding experimental values and fundamental
vibrational frequencies within 5 cm$^{-1}$, or even 1 cm$^{-1}$ in some cases
\citep{Huang11, Fortenberry11HOCO, Fortenberry11cHOCO, Fortenberry12hococat,
Huang13NNOH+}, rotational constants, dipole moments, spectroscopic constants,
and fundamental vibrational frequencies are provided for these two open-shell
noble gas cations and their myriad isotopologues.

\section{Computational Details}

The highly-accurate structural and spectroscopic data necessary to examine the
presence of NeH$_2$$^+$ and ArH$_2$$^+$ in any reaction scheme is derived here
from quartic force fields (QFFs).  QFFs are fourth-order Taylor series
approximations to the nuclear Hamiltonian \citep{Fortenberry13Morse} and are of
the form:
\begin{equation}
V=\frac{1}{2}\sum_{ij}F_{ij}\Delta_i\Delta_j +
\frac{1}{6}\sum_{ijk}F_{ikj}\Delta_i\Delta_j\Delta_k +
\frac{1}{24}\sum_{ijkl}F_{ikjl}\Delta_i\Delta_j\Delta_k\Delta_l, 
\label{VVib}
\end{equation} 
where $\Delta_i$ are displacements and $F_{ij\ldots}$ are the force constants.
These methods provide low-cost (relative to producing a global PES) but still
high-accuracy computationally-derived spectroscopic data.  The most accurate
QFFs \citep{Huang08, Huang09, Huang11}  begin from coupled cluster singles,
doubles, and perturbative triples [CCSD(T)] \citep{Rag89} geometry
optimizations with the cc-pV5Z basis sets \citep{Dunning89, cc-pVXZ}.  Ar,
however, requires the cc-pV(5+d)Z basis set \citep{cc-pVXZ}, but this will
simply be called cc-pV5Z (and likewise for similar basis sets) for ease of
discussion.  The geometrical parameters are further corrected for the
correlation of the core electrons by CCSD(T) geometry optimizations with the
Martin-Taylor (MT) core-correlating basis set \citep{Martin94} derived to treat
these properties.  In this case, the difference between the CCSD(T)/MT bond
lengths with core and those without core electrons are added to the
CCSD(T)/cc-pV5Z bond lengths.  All computations are based on restricted
open-shell reference wavefunctions \citep{Gauss-ROHF91, Lauderdale-ROHF91,
Watts-ROHF93} for these open-shell cations.

From this linear reference geometry, a grid of 55 symmetry-unique points are
used to define the QFF.  Displacements of the two bond lengths are made in
$\Delta_p$ increments of 0.005 \AA, and the bond angle $\Delta_p$ increments
are 0.005 radians.  The Ng$-$H bond length is coordinate 1, the H$-$H bond
length is coordinate 2, and the Ng$-$H$-$H bond angle is coordinate 3, where
this latter term is treated in a linear-bending term defined in the INTDER
program \citep{intder} and discussed elsewhere \citep{Huang11, Huang13C3H+}.
Degenerate with and perpendicular to coordinate 3 is the other Ng$-$H$-$H bond
angle, which is designated as coordinate 4.  However, the symmetry relationship
of this coordinate to coordinate 3 negated explicit treatment of this
angle \citep{Huang09, Huang13C3H+}.  At each point, CCSD(T) energies are computed
with the aug-cc-pVTZ, aug-cc-pVQZ, and aug-cc-pV5Z basis sets \citep{Dunning89,
aug-cc-pVXZ}, which are extrapolated to the one-particle complete basis set
(CBS) limit via a three-point formula \citep{Martin96}.  To these CBS energies,
corrections for core correlation from the MT basis set with and without core
electrons and scalar relativistic \citep{Douglas74} corrections are also made.
This definition of the QFF is called the CcCR QFF from the ``C'' for CBS,
``cC'' for core correlation, and ``R'' for relativity \citep{Fortenberry11HOCO}.

The CcCR force constants are derived from a fitting of the QFF via a
least-squares approach.  This also produces the equilibrium geometry with a sum
of squared residuals on the order of 10$^{-19}$ a.u.$^2$ for NeH$_2$$^+$ and
10$^{-15}$ a.u.$^2$ for ArH$_2$$^+$.  Utilizing this new minimum, the force
constants are refit to produce zero gradients and positive quadratic terms
along with subsequently more complete cubic and quartic terms.  The force
constants are then incorporated in the linear nuclear Hamiltonian via the
SPECTRO program \citep{spectro91}, where second-order perturbation theory
produces the spectroscopic constants \citep{Mills72, Watson77} and vibrational
frequencies \citep{Papousek82}.  Most of the electronic structure computations
make use of the PSI4 quantum chemistry package \citep{psi4}, but the scalar
relativistic energy points as well as the CCSD(T)/aug-cc-pV5Z dipole moment
computations utilize MOLPRO2012.1 \citep{MOLPRO2012} made available on the
Pittsburgh Supercomputing Center's Blacklight Cluster.

\section{Results and Discussion}

\subsection{NeH$_2$$^+$ Structural and Spectroscopic Considerations}

The force constants for $^2\Sigma^+$ NeH$_2$$^+$ are given in Table \ref{Nefc}
following the coordinate ordering defined in the previous section.  This linear
structure is clearly a minimum since all of the second derivatives are
positive, indicating an upward curve to the PES.  The structural and
spectroscopic data for $^{20}$NeH$_2$$^+$ and its deuterated isotopologues are
given in Table \ref{Ne20} while the geometric and spectroscopic data for the
less abundant neon isotopic cation, $^{22}$NeH$_2$$^+$, and its deuterated
isotopologues are in Table \ref{Ne22}.  The force constants remain the same for
any of the isotopologues since they are computed within the Born-Oppenheimer
approximation.  The NeH$_2$$^+$ and NeHD$^+$ molecules, regardless of neon
isotope, require input of a $2\nu_2=\nu_1$ type-1 Fermi resonance as well as a
$\nu_2+\nu_1=\nu_2$ type-2 Fermi resonance.  Likewise, the NeDH$^+$ and
NeD$_2$$^+$ systems require an additional $2\nu_3=\nu_2$ type-1 Fermi resonance
as well as a further $\nu_2+\nu_3=\nu_3$ type-2 Fermi resonance.

In the course of this study, it was found that the quartic terms for the $\pi$
bending mode are unreliable due to the flatness of the PES in this coordinate
coupled with only having one, with at most two, significant figures in the
quartic force constants.  As such, these terms are dropped from the definition
of the PES giving a QFF for only the totally-symmetric modes (the two
stretches) and a cubic force field PES for the pair of degenerate bending
modes.  Even so, all of the CcCR spectroscopic constants of the ground vibrational
state are produced in the same manner and, subsequently, the same accuracy as
those produced in the previous studies \citep{Huang11, Fortenberry11HOCO,
Fortenberry11cHOCO, Fortenberry12hococat, Huang13NNOH+}, since only cubic terms
are necessary for the computation of $B_0$ and the other desired spectroscopic
constants in the first place \citep{spectro91}.  Additionally, the $\nu_1$ and
$\nu_2$ frequencies, as well as their vibrationally-averaged rotational
constants, are defined and computed in the same procedure as has been previously
benchmarked for high accuracy.  The $\nu_3$ frequency may not be as reliable
since its largest contributor to the Taylor series definition of the potential
is cubic, but this is still an improvement over simple harmonic frequencies and
will give some indication as to the rotational structure of this vibrational
state.

The vibrational frequencies for this noble gas compound are very different from
``standard'' atoms.  For instance, the $\nu_1$ H$-$H stretch is over half the
value of what is known for the stretching motion of hydrogen atom at more than
4000 cm$^{-1}$ or less than 2.5 microns \citep{Irikura07}, but actually quite
close to the roughly 2000 cm$^{-1}$ (5.0 microns) fundamental vibrational
frequency of H$_2$$^+$ \citep{Herzberg72, Roth08}. As a result of these closely
coincidental values, it is interpreted that the presence of the H$-$H bond
behaves much like it does in the hydrogen molecular cation, but the neon atom
affects this motion to a small degree indicating that some bonding is present.
The $\nu_2$ Ne$-$H stretch fundamental is substantially lower than most other
heavy atom to hydrogen stretches, which are typically in the 3300 to 3600
cm$^{-1}$ (2.8 $-$ 3.0 micron) region, but the harmonic frequency is many times
higher than van der Waals Ne$-$X stretching frequencies \citep{Grandinetti11}
indicating that the Ne$-$H bond itself is also weak but certainly functional.
Finally, the bending motion is also lower than it would be in most purely
``covalent'' structures.  Hence, the vibrations of NeH$_2$$^+$ imply that this
molecule is weakly associated but has some level of electron sharing in its
stabilization making it more than merely a van der Waals complex. 



The bond lengths for NeH$_2$$^+$ further indicate that this molecule is more
than a van der Waals complex.  The Ne$-$H bond is longer than other hydrogen
bonds of the same period, but much shorter than would be found in purely
electrostatic interactions between noble gases and ligands
\citep{Grandinetti11}. The equilibrium bond lengths remain the same upon
isotopomerization since the potential has been formulated, again, within the
Born-Oppenheimer approximation.  Similarly, the dipole moment remains unchanged
for our set of computations.  The CCSD(T)/aug-cc-pV5Z dipole moment computed
from the centre-of-mass at the origin is very high here at 5.61 D making this
cation very rotationally bright regardless of the isotopologue examined.  The
$R_{\alpha}$ vibrationally-averaged bond lengths are affected by the
differences in the various isotopic masses as shown in Tables \ref{Ne20} and
\ref{Ne22} which, in turn, influence the $B$-type rotational constants.
Inclusion of the $^{22}$Ne over $^{20}$Ne lowers $B_0$ by 554.0 MHz, which will
produce noticeably different rotational spectra at high enough resolution, but
the near doubling of the hydrogen masses upon deuteration will have a much
greater effect on any observed rotational lines.  Also from the tables, the
vibrational modes are all affected by isotopic shifts that further affects the
$B_1$, $B_2$, and $B_3$ constants giving new rotational structure for the
vibrationally excited states.

\subsection{ArH$_2$$^+$ Structural and Spectroscopic Considerations}

The force constants for $^2\Sigma^+$ ArH$_2$$^+$ are given in Table \ref{Arfc}
including, again, all positive second derivatives.  Argon has three stable
isotopes: $^{36}$Ar, $^{38}$Ar, and $^{40}$Ar.  $^{36}$ArH$^+$ was detected in
the Crab nebula \citep{Barlow13}, even though $^{40}$Ar is the more abundant
isotope on Earth.  The reasons for the difference in abundances of the two
isotopes have been well documented \citep{Cueto14}, but inclusion of all three
forms of argon is warranted in this study for the sake of completeness.  The
structural, spectroscopic, and vibrational values for the twelve different
isotopologues (including deuteration) are given in Tables \ref{Ar36},
\ref{Ar38}, and \ref{Ar40}.

The resonances required for the SPECTRO computations are not affected by
inclusion of the different argon isotopes.  However, the deuteration affects
these necessary inputs since it influences the rovibrational structure of the
molecules more noticeably.  No resonances of any kind are needed for
ArH$_2$$^+$.  The resonances required for ArHD$^+$ include a $2\nu_2=\nu_1$
type-1 Fermi resonance, a $\nu_2+\nu_1=\nu_2$ type-2 Fermi resonance, and
$\nu_2 / \nu_3$ Coriolis resonances.  ArDH$^+$ is reliant upon $2\nu_3=\nu_1$
and $2\nu_3=\nu_2$ type-1 Fermi resonances, a $\nu_2+\nu_3=\nu_3$ type-2 Fermi
resonance, and a $\nu_2 / \nu_1$ Darling-Dennison resonance.  The ArD$_2$$^+$
resonances are the same as those for ArDH$^+$ but without the Darling-Dennison
resonance.  A similar issue with the quartic terms of the $\pi$-bending
coordinate were also encountered with ArH$_2$$^+$ leading once again to the use
of the CcCR QFF for the totally-symmetric terms but only the CcCR cubic force
field for the bending mode(s).

The Born-Oppenheimer dipole moment computed from the centre-of-mass for
ArH$_2$$^+$ is not as high as it is for the neon compound but still quite
significant at 4.37 D.  The Ar$-$H bond length is around 1.44 \AA\ for each
isotopologue, which is about 0.1 \AA\ longer than another set of third-row atom
containing compounds, $cis$- and $trans$-HSCO, where the S$-$H bond lengths are
1.34-1.35 \AA\ \citep{Fortenberry14HOCS}.  Even so, it is still in the range of
fairly well bonded hydrogen atoms for larger atoms of this size and cannot be
interpreted as a van der Waals complex, even though the equilibrium H$-$H bond
length of 1.110 \AA\ is very close to the 1.053 \AA\ bond length in
H$_2$$^+$ \citep{Herzberg72}.

The anharmonic vibrational frequencies for the ArH$_2$$^+$ isotopologues are
significantly less than their neon-containing counterparts, as one would expect
for inclusion of a heavier atom$-$argon in this case.  However, the decrease in
the H$-$H stretch from 1801.5 cm$^{-1}$ (5.551 microns) in the standard
isotopologue of NeH$_2$$^+$ to 1229.8 cm$^{-1}$ (8.131 microns) in
$^{36}$ArH$_2$$^+$ indicates that this bond is even more perturbed by the
presence of the argon than the neon when compared to mere H$_2$$^+$.  The
$\nu_2$ Ar$-$H stretch is 201.7 cm$^{-1}$ less at 590.5 cm$^{-1}$ (16.93
microns) than the $^{20}$NeH$_2$$^+$ $\nu_2$ Ne$-$H stretch.  However, this
mode is strongly anharmonic for ArH$_2$$^+$, where the anharmonicities are on
the order of 500 cm$^{-1}$.  Single deuteration of the terminal hydrogen
significantly reduces the anharmonicities, on the order of a half, and the
ArDH$^+$ $\nu_2$ stretches are almost harmonic in comparison.  Also, the
$\nu_2$ Ar$-$H stretch is significantly lower in ArH$_2$$^+$ than it is in
ArH$^+$ where the frequency is 2592.65 cm$^{-1}$ or 3.857 microns
\citep{Cueto14} indicating that the Ar$-$H bond is stronger in the diatomic
cation than it is in ArH$_2$$^+$.  However, the 590.5 cm$^{-1}$ $\nu_2$ Ar$-$H
stretch is notably higher than the comparable 350 cm$^{-1}$ Ar$-$H$_2$ stretch
observed for the ArH$_2$ neutral van der Waals complex \citep{McKellar96}.  The
exception to the decrease in frequency for the ArH$_2$$^+$ fundamentals as
compared to NeH$_2$$^+$ is the $\pi$-bending mode.  For the standard
isotopologue, the NeH$_2$$^+$ bending mode is 98.5 cm$^{-1}$ less than its
$^{36}$ArH$_2$$^+$ counterpart.  However, again, the lack of quartic terms in
these descriptions does not lend strong credence to any conclusions taken about
these modes besides a general location as to where these bending frequencies
can be found.

The vibrational frequencies indicate that ArH$_2$$^+$ is not quite as strongly
bound as ArH$^+$ but is more strongly associated than neutral ArH$_2$.  This is
corroborated by the force constants, which, according to Badger's Law, indicate
bond strength since it is proportional to the force constant of the vibration.
From Table \ref{Arfc}, the $F_{11}$ force constant corresponds to the Ar$-$H
stretch and is 1.895 279 mdyn/\AA.  Using the harmonic approximation from the
vibrational frequency derived by \citet{Cueto14} (2592.65
cm$^{-1}$), the ArH$^+$ force constant is 3.883 866 mdyn/\AA, nearly double
that in ArH$_2$$^+$.  Concurrently, the $F_{22}$ force constant is 0.700 608
mdyn/\AA\ indicating that the H$-$H bond is half as strong as it is in
H$_2$$^+$, where the harmonic force constant is 1.425 879 mdyn/\AA\ from a
2191.13 cm$^{-1}$ frequency \citep{Roth08}. Regardless of the strength of the
bonds and the degree to which electrons are shared between the Ar atom and the
H$_2$$^+$ cation, the rotational constants and other spectroscopic data for
ArH$_2$$^+$ as provided in Tables \ref{Ar36}, \ref{Ar38}, and \ref{Ar40} can be
applied to spectral analysis of the ISM or simulated laboratory experiments.

\subsection{Astrochemical Pathways and Spectra}

The main question regarding the detectability of these compounds in the ISM or
their presence in interstellar chemical pathways is whether or not these
structures are stable and, if so, by how much.  Since zero gradients are
determined along with positive second derivates (See Tables \ref{Nefc} and
\ref{Arfc}) these structures are minima on their respective potential surfaces.
In fact, these $^2\Sigma^+$ linear structures are the global minima on their
respective PESs.  In Table \ref{Diss}, the major possible dissociation pathways
are listed for each noble gas molecule with energies provided relative to each
triatomic minimum.  According to ``gold standard'' \citep{Helgaker04}
CCSD(T)/aug-cc-pVTZ computations, the most likely direct dissociation pathway
for NeH$_2$$^+$ is into the Ne atom and the hydrogen molecular cation.  The
NeH$_2$$^+$ minimum sits 0.560 eV below these two products.  All the
fundamental vibrational frequencies are below this value showing that
vibrational transitions, in addition to pure rotational spectra, can be
detected for this system if the conditions are correct for its interstellar
formation.  The next-lowest dissociation pathway is for NeH$_2$$^+$ to lose a
hydrogen atom to form NeH$^+$, but this costs nearly twice as much energy at
1.031 eV.  This closely mirrors the previous Ne + H$_2$$^+ \rightarrow$ NeH$^+$
+ H PES computed by \cite{Gamallo13} and shown in Figure 1 therein.

The energetics of these two pathways are reversed for ArH$_2$$^+$, which is the
global minimum on its PES.  The lowest energy dissociation pathway is, again
from Table \ref{Diss}, into ArH$^+$ and a hydrogen atom where only 0.493 eV is
required to break the triatomic noble gas cation down.  The dissociation into
the Ar atom and hydrogen molecular cation is much higher at 1.838 eV and
further breakdown into Ar + H + H$^+$ requires another 2.429 eV, comparable to
2.634 eV predicted by \cite{Song03} with a lower level of theory.  Since the
most likely way for ArH$_2$$^+$ to dissociate, and do so by a wide margain, is
to form ArH$^+$ and a hydrogen atom, it seems likely that ArH$_2$$^+$ is a
precursor to the formation of ArH$^+$ in the ArH$^+$ + H$_2$ PES.  Since this
dissociation energy corresponds to an equivalent temperature of 5725.7 K, hot
regions like the Crab nebula where ArH$^+$ has been detected should see
ArH$_2$$^+$ break down fairly easily into the previously detected ArH$^+$
diatomic cation.  Cooler regions, such as molecular clouds, should retain a
fair amount of ArH$_2$$^+$ before dissociation into ArH$^+$.

\begin{figure}
\caption{Visual depiction for the relative energies of the ArH$_2$$^+$ PES.}
\includegraphics[width = 3.3 in]{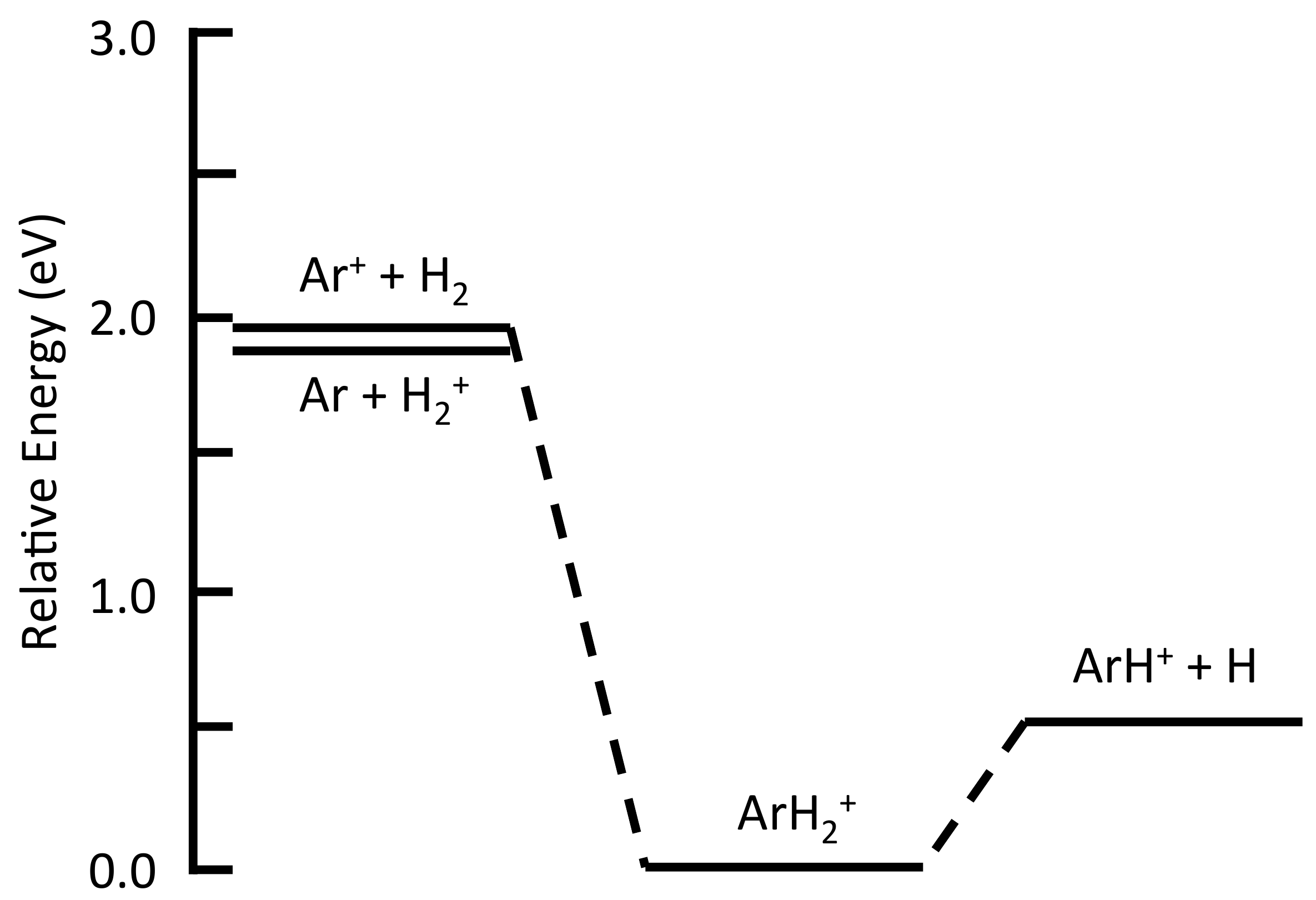}
\label{ArPES}
\end{figure}

\begin{figure}
\caption{Visual depiction for the relative energies of the NeH$_2$$^+$ PES.}
\includegraphics[width = 3.3 in]{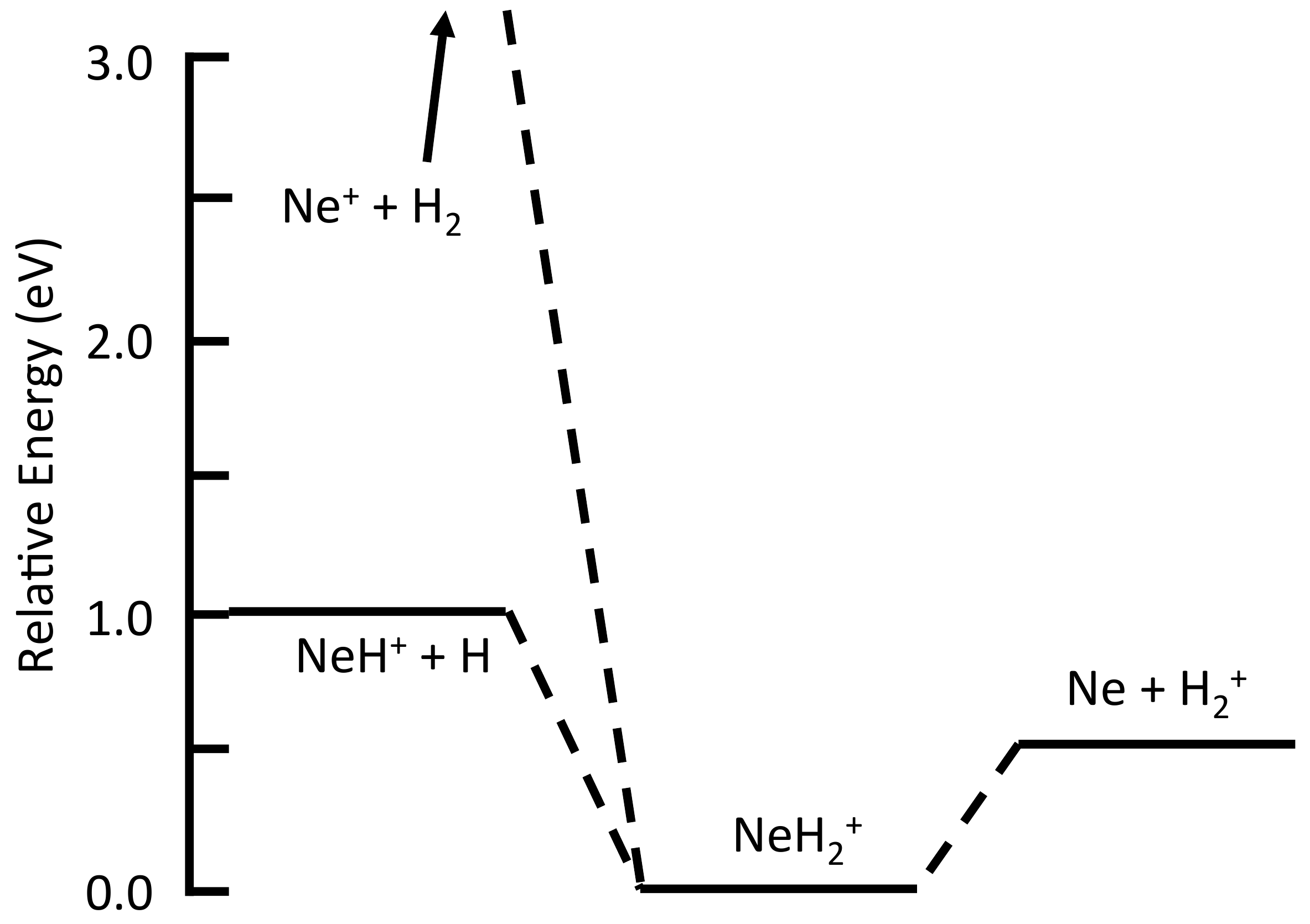}
\label{NePES}
\end{figure}

Therefore, it is plausible that ArH$^+$ forms, at least in part, from
ArH$_2$$^+$ in the ISM.  It may be that ArH$_2$$^+$ is the necessary stable or
semi-stable intermediate in the Ar$^+ + \mathrm{H}_2 \rightarrow \mathrm{ArH}^+
+ \mathrm{H}$ mechanism as proposed by \citet{Barlow13} since formation of
ArH$_2$$^+$ from Ar$^+ + \mathrm{H}_2$ produces 1.930 eV (see Table \ref{Diss})
of energy whereas direct, bimolecular dissociation produces 1.436 eV.  Granted,
0.493 eV must be imparted into the system to break ArH$_2$$^+$ down into ArH$^+
+ \mathrm{H}$, but this stabilization brought about by the presence of
ArH$_2$$^+$ in the reaction scheme may speed up the formation of ArH$^+$ in the
ISM.  Additionally, the role of ArH$_2$$^+$ may also be an intermediate in the
alternative $\mathrm{Ar} + \mathrm{H}_2^+ \rightarrow \mathrm{ArH}^+ +
\mathrm{H}$ mechanism, as well, which is slightly more energetically favorable
at 1.838 eV above the ArH$_2$$^+$ minimum.  These possible reactions schemes
are depicted in Figure \ref{ArPES}.  Additionally, ArH$_2$$^+$ may even be a
simple degradation precursor to ArH$^+$ in cool environments after collision
with another body.  Regardless, ArH$_2$$^+$ should be detectable, even if in
small amounts as it would be functioning as an intermediate, due to its large
dipole moment.  Finally, NeH$^+$ will probably not be as abundant as ArH$^+$ as
has been shown recently \citep{Schilke14}. The most probable dissociation
pathway for NeH$_2$$^+$ is into Ne and H$_2$$^+$, as depicted in Figure
\ref{NePES} corroborating previous work on this PES \citep{Gamallo13}, making
any potential NeH$^+$ detection unlikely.  This is especially true if the Ne$^+
+ \mathrm{H}_2$ reaction is necessary to create NeH$^+$ as it has been
suggested for Ar$^+ + \mathrm{H}_2$ to form ArH$^+$ \citep{Barlow13, Schilke14,
Roueff14}.

If detected as a stable species in its own right, ArH$_2$$^+$ could also serve
as a contributing factor indicating the temperature of a given warm
astronomical environment since it should dissociate around 5700 K.  Even though
its most likely dissociation energy is at an equivalent temperature of 6500 K,
NeH$_2$$^+$ is not as flexible for use as a thermometer.  Again, its primary
dissociation pathway is into the neutral noble gas atom and H$_2$$^+$.  No
rotationally bright product is created that could be interpreted as unique to
NeH$_2$$^+$ dissociation as it is in ArH$_2$$^+ \rightarrow \mathrm{ArH}^+ +
\mathrm{H}$.


\begin{figure*}
\caption{The rotational spectrum of the ground and excited vibrational states
of ArH$_2$$^+$ in THz at 40 K with the shorter lines corresponding to
$^{38}$ArH$_2$$^+$ and the taller lines to $^{36}$ArH$_2$$^+$ based on the
relative natural abundances of the argon isotopes.}
\includegraphics[width = 6.5 in]{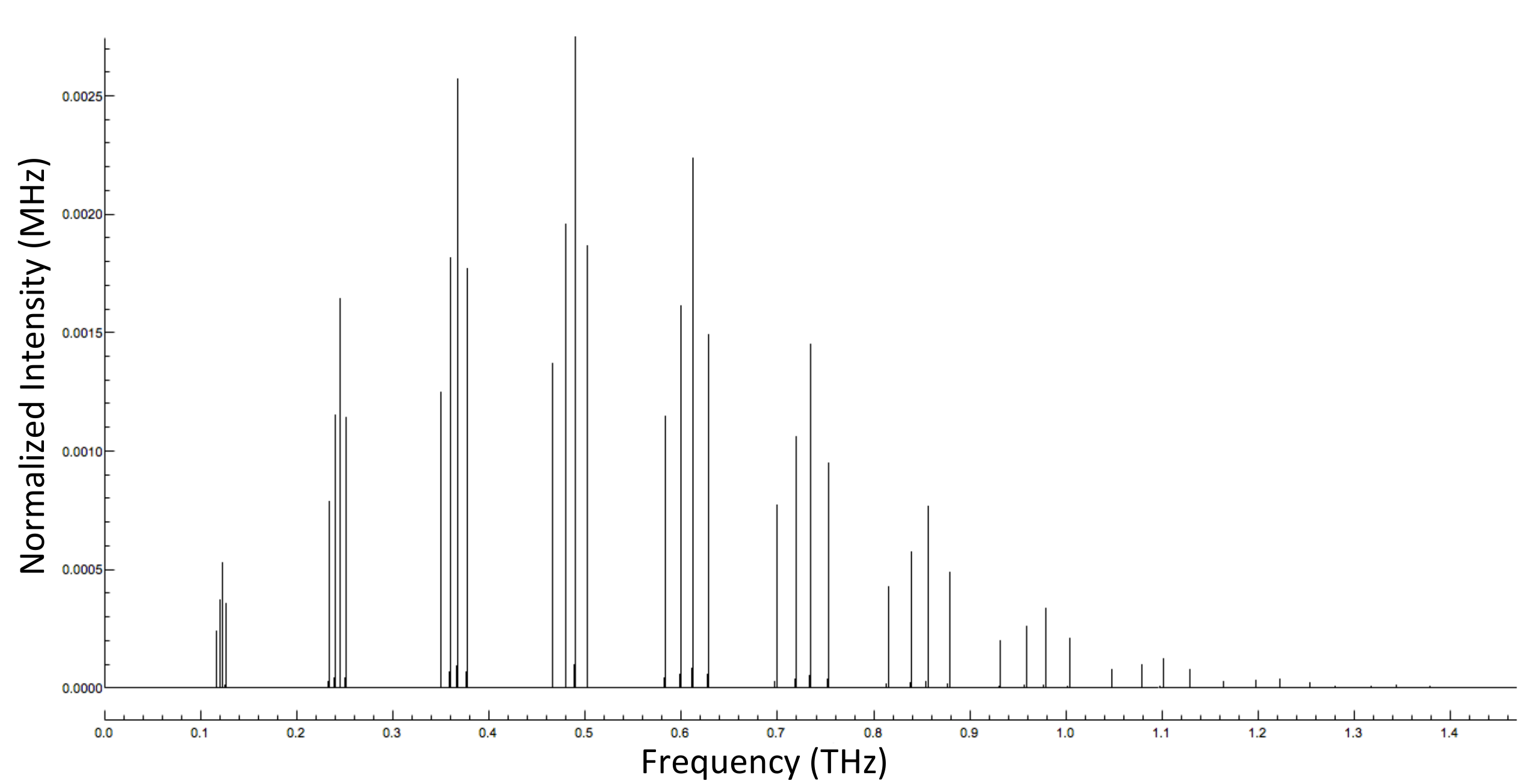}
\label{ArRot40}
\end{figure*}

The rotational spectra of NeH$_2$$^+$ and ArH$_2$$^+$ can be produced from the
spectroscopic data provided herein.  Due to its potential relationship of
ArH$_2$$^+$ to ArH$^+$, the rotational spectrum of NeH$_2$$^+$ is left to the
reader to deduce.  The natural $[^{36} \mathrm{Ar} / ^{38} \mathrm{Ar}]$ ratio
of roughly 5.25 should produce a sizable intensity for the rotational
transitions of the heavier isotopologue of ArH$_2$$^+$ in addition to the
stronger $^{36}$ArH$_2$$^+$ rotational lines.  Figure \ref{ArRot40}
demonstrates what the full rotational spectrum for ArH$_2$$^+$ should look like
at 40 K with the intensities of $^{36}$ArH$_2$$^+$ normalized and the
intensities of $^{38}$ArH$_2$$^+$ scaled from the permanent dipole moment and
the relative natural abudnance of $^{38}$Ar.  The simple model linear rotor for
inclusion of up to sextic distortion,
\begin{equation}
\Delta E(J+1 \rightarrow J) = 2B(J+1) - 4D(J+1)^3 +H(J+1)^3 [(J+2)^3 - J^3],  
\label{Ej}
\end{equation}
is employed here within the PGOPHER \citep{pgopher} program making use of the
spectroscopic constants given in Tables \ref{Ar36} and \ref{Ar38}.  Transitions
up to the $J = 10 \rightarrow 9$ line located just around 1.2 THz should be
visible, but the $J = 4 \rightarrow 3$ line appears to be the brightest at
these temperatures.

\begin{figure*}
\caption{The rotational spectrum of ArH$_2$$^+$ in THz at 5000 K where the
vibrationally excited rotational lines of $^{36}$ArH$_2$$^+$ and
$^{38}$ArH$_2$$^+$ are visible.}
\includegraphics[width = 6.5 in]{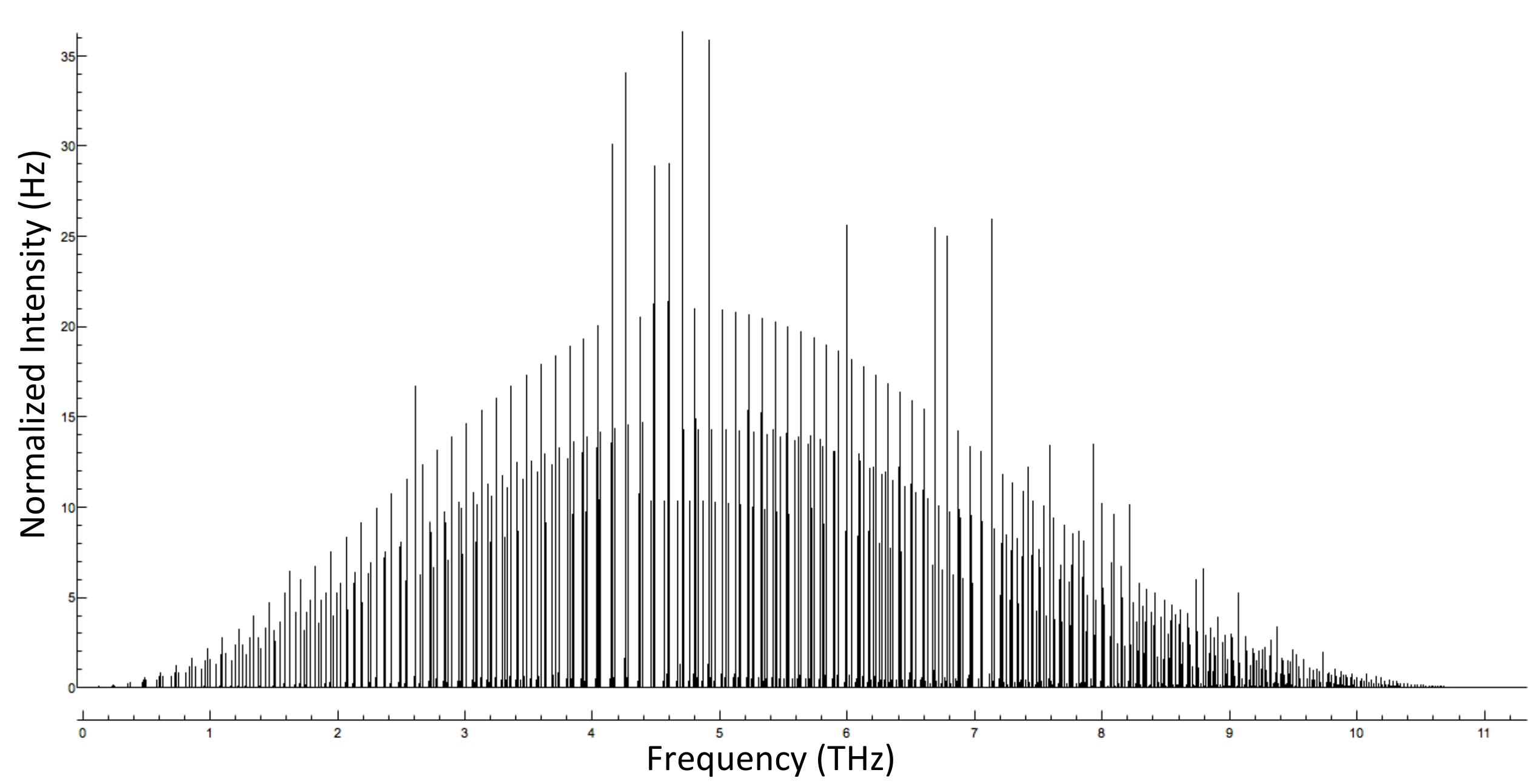}
\label{ArRot5000}
\end{figure*}

At temperatures closer to that found in the Crab nebula or similar
environments, the spectrum changes markedly, as depicted in Figure
\ref{ArRot5000} for 5000 K.  In addition to the increase in the number of pure
rotational lines present, the vibrationally excited states are substantially
more populated at this temperature with Boltzmann proportions of 0.71, 0.84,
and 0.83, respective of $\nu_1$, $\nu_2$, and $\nu_3$, normalized to the pure
rotational transitions.  As such, the spectrum becomes more complicated in this
idealized depiction.  The $J = 42 \rightarrow 41$ transition is the brightest
at this temperature and should be found in the 4.9 THz range.  The spin-orbit
peak splitting present in such radicals is not modeled here due to limitations
in our methodology but would be a further factor in the interstellar detection
of these noble gas-containing triatomic cations regardless of the environmental
temperatures.

To aid in experimental characterization, the pure rotational line list for
$\Delta E(J+1 \rightarrow J), J \leq 9,$ of the most abundant isotopes of
argon, $^{36}$Ar and $^{38}$Ar, is given in Table \ref{LLL} to complement
Figures \ref{ArRot40} and \ref{ArRot5000}.   Experience has shown that the
rotational constants may vary from experiment by 10-25 MHz.  As such the
variance from experiment for the lines listed here will grow for higher $J$,
but the relative transition frequencies and the relative intensities will, no
doubt, benefit searches for this noble gas cation.

\section{Conclusions}

The rotational and rovibrational spectroscopic data provided here should assist
in the potential detection of NeH$_2$$^+$ and ArH$_2$$^+$ in the ISM.  The pure
rotational and $\nu_1$/$\nu_2$ data are computed in the same CcCR QFF fashion as
has been done previously to produce highly-accurate results.  Since the
$\pi$-bending mode could not fully employ this same approach, the spectroscopic
data generated by the cubic force field will certainly provide deeper insight
into the vibrational and rovibrational nature of the $\nu_3$ modes than
harmonically derived values can.  The bond lengths and vibrational frequencies
show clear demarkation from simple van der Waals complexes for these two noble
gas cations.  NeH$_2$$^+$ and ArH$_2$$^+$ are more covalent in nature but
cannot be considered purely covalent.  Both species exhibit large dipole
moments that should make them more easily detected than other ``trace'' species
in the ISM.

Additionally, the different lowest-energy CCSD(T)/aug-cc-pVTZ dissociation
pathways of NeH$_2$$^+$ into Ne and H$_2$$^+$ with ArH$_2$$^+$ into ArH$^+$ and
H showcase the differences in the two noble gas atoms.  It is therefore likely
that ArH$_2$$^+$ is a precursor to ArH$^+$ as a reaction intermediate between
Ar + H$_2$$^+$ or Ar$^+ + \mathrm{H}_2$  where ArH$^+$ and a hydrogen atom are
the major products.  In either case, ArH$_2$$^+$ is the global minimum on the
PES.  If ArH$_2$$^+$ is involved in the formation of ArH$^+$, it is unlikely,
then, that NeH$^+$ will form in the ISM in a similar fashion since this is not
the most energetically favorable pathway.  As a result, detection of the neon
hydride cation is doubtful as recent surveys have shown \citep{Schilke14}.
Regardless, the reference data provided here, including the rotational spectra,
should assist in analysis of the noble gas chemistry of the Crab nebula where
ArH$^+$ has been detected or in other regions in which other noble gas
compounds may be observed.

\section{Acknowledgements}

The authors are grateful to Georgia Southern University for providing start-up
funds that supported this research in the form of student salaries and the
purchase of computer equipment.  The Georgia Southern College of Science and
Mathematics also provided support for WJM and RCF from an ``Interdisciplinary
Pilot Grant''.  This work used the Extreme Science and Engineering Discovery
Environment (XSEDE), which is supported by National Science Foundation grant
number ACI-1053575.  The authors would also like to thank Jim LoBue of Georgia
Southern for providing many of the references cited as well as Mallory L.~Theis
and Lauren F.~Fortenberry for assistance in editing the manuscript.
Prof.~Jonathan Tennyson of the University College London is gratefully thanked
for pointing out a systemic error in our original spectra that has since been
corrected.  The reviewer is also acknowledged for the very useful insights
provided throughout the review process.

\bibliographystyle{mn2e}







\renewcommand{\baselinestretch}{1}
\begingroup
\begin{table}

\caption{The NeH$_2$$^+$ CcCR Simple-Internal Force Constants (in
mdyn/\AA$^n$$\cdot$rad$^m$)$^a$.}

\label{Nefc}

\centering

\begin{tabular}{c r c r c r c r c r}
\hline

F$_{11}$ & 1.039 234 & F$_{331}$ & -0.1783 & F$_{3311}$ & 0.13 \\ 
F$_{21}$ & 0.383 519 & F$_{332}$ & -0.0352 & F$_{3321}$ & 0.39 \\  
F$_{22}$ & 1.197 493 & F$_{441}$ & -0.1783 & F$_{3322}$ & 0.37 \\  
F$_{33}$ & 0.061 686 & F$_{442}$ & -0.0352 & F$_{3333}$ & -1.72 \\ 
F$_{44}$ & 0.061 686 & F$_{1111}$ & 52.85 & F$_{4411}$ & 0.13 \\  
F$_{111}$ & -7.9408 & F$_{2111}$ & 5.13 & F$_{4421}$ & 0.39 \\ 
F$_{211}$ & -1.7665 & F$_{2211}$ & 2.45 & F$_{4422}$ & 0.37 \\ 
F$_{221}$ & -0.0664 & F$_{2221}$ & -0.59 & F$_{4433}$ & -0.49 \\
F$_{222}$ & -5.7851 & F$_{2222}$ & 25.68 & F$_{4444}$ & -1.72 \\
 
\hline
\end{tabular}

$^a$1 mdyn $=$ $10^{-8}$ N; $n$ and $m$ are exponents corresponding to the
number of units from the type of modes present in the specific force constant.
\AA\ are fitting for modes of bond stretches and degrees for the degenerate
bending modes.
 
\end{table}
\endgroup

\renewcommand{\baselinestretch}{1}
\begingroup
\begin{table*}

\caption{The CcCR Zero-Point ($R_{\alpha}$ vibrationally-averaged) and
Equilibrium Structures, Rotational Constants, CCSD(T)/aug-cc-pV5Z Dipole
Moment, Vibration-Rotation Interaction Constants, and Quartic (D, $\tau$) and
Sextic (H) Distortion Constants of $^{20}$NeH$_2$$^+$ and Deuterated
Isotopologues.}

\label{Ne20}

\centering

\small

\begin{tabular}{l | r r r r} 
\hline\hline

                   & NeH$_2$$^+$   & NeHD$^+$      & NeDH$^+$       & NeD$_2$$^+$   \\ 
\hline
r$_0$(Ne$-$H$_1$)  & 1.234 314 \AA & 1.231 167 \AA & 1.230 908 \AA  & 1.228 534 \AA  \\
r$_0$(H$_1-$H$_2$) & 1.105 287 \AA & 1.105 987 \AA & 1.101 895 \AA  & 1.104 149 \AA  \\
$B_0$              & 78 699.8 MHz  & 46 203.8 MHz  & 67 591.3 MHz   & 42 820.6 MHz  \\
$\alpha^B$ 1       &  2406.6 MHz   &  2444.3 MHz   &  36.8 MHz      &  949.2 MHz      \\
$\alpha^B$ 2       &  3745.6 MHz   &  1329.0 MHz   &  3324.5 MHz    &  1474.6 MHz      \\
$\alpha^B$ 3       & -1494.2 MHz   & -1066.2 MHz   & -573.0 MHz     & -572.8 MHz      \\
$\tau_{aaaa}$      & -9.417 MHz    & -3.037 MHz    & -7.795 MHz     & -2.735 MHz    \\
\hline                                                    
r$_e$(Ne$-$H$_1$)  & 1.211 598 \AA & --            & --             & --             \\
r$_e$(H$_1-$H$_2$) & 1.100 735 \AA & --            & --             & --             \\
$B_e$              & 80 281.7 MHz  & 47 024.2 MHz  & 68 698.9 MHz   & 43 459.7 MHz  \\
$D_e$              &  2.354 MHz    &  0.759 MHz    &  1.949 MHz     &  0.684 MHz      \\
$H_e$              &-49.825  Hz    & -3.043 Hz     & -63.774  Hz    & -7.235  Hz      \\
$\mu_z$$^a$        &  5.61 D       & --            & --             & --             \\ 
\hline                                                    
$\omega_1(\sigma)$ H$-$H   & 1907.4 cm$^{-1}$ & 1722.9 cm$^{-1}$ & 1634.8 cm$^{-1}$ & 1349.8 cm$^{-1}$ \\
$\omega_2(\sigma)$ Ne$-$H  &  970.4 cm$^{-1}$ &  777.1 cm$^{-1}$ &  819.0 cm$^{-1}$ &  716.9 cm$^{-1}$ \\
$\omega_3(\pi)$ Bend       &  633.7 cm$^{-1}$ &  599.0 cm$^{-1}$ &  495.6 cm$^{-1}$ &  450.3 cm$^{-1}$ \\
Harmonic Zero-Point        & 2072.6 cm$^{-1}$ & 1849.0 cm$^{-1}$ & 1722.5 cm$^{-1}$ & 1483.7 cm$^{-1}$\\
$\nu_1(\sigma)$ H$-$H      & 1801.5 cm$^{-1}$ & 1611.3 cm$^{-1}$ & 1526.2 cm$^{-1}$ & 1323.3 cm$^{-1}$ \\
$\nu_2(\sigma)$ Ne$-$H     &  792.2 cm$^{-1}$ &  636.6 cm$^{-1}$ &  723.2 cm$^{-1}$ &  616.0 cm$^{-1}$ \\
$\nu_3(\pi)$ Bend          &  557.1 cm$^{-1}$ &  530.3 cm$^{-1}$ &  448.6 cm$^{-1}$ &  404.8 cm$^{-1}$ \\
Zero-Point                 & 2002.0 cm$^{-1}$ & 1794.0 cm$^{-1}$ & 1672.8 cm$^{-1}$ & 1448.5 cm$^{-1}$ \\
\hline                                                    
$B_1$       & 76 293.2 MHz  & 43 759.50 MHz  & 67 554.50 MHz   & 41 871.40 MHz  \\
$B_2$       & 74 954.2 MHz  & 44 874.80 MHz  & 64 266.80 MHz   & 41 346.00 MHz  \\
$B_3$       & 80 194.0 MHz  & 46 778.40 MHz  & 68 164.30 MHz   & 43 393.30 MHz  \\

\hline
\end{tabular}

$^a$The NeH$_2$$^+$ coordinates (in \AA\ with the centre-of-mass at the origin)
used to generate the Born-Oppenheimer dipole moment component are: Ne, 0.000000,
0.000000, -0.161373; H$_1$, 0.000000, 0.000000, 1.050226; H$_2$, 0.000000,
0.000000, 2.150962\\

\end{table*}
\endgroup

\renewcommand{\baselinestretch}{1}
\begingroup
\begin{table*}

\caption{The CcCR $^{22}$NeH$_2$$^+$ and Deuterated Isotopologues Spectroscopic
Data.}

\label{Ne22}

\centering

\small

\begin{tabular}{l | r r r r} 
\hline\hline

                   & NeH$_2$$^+$   & NeHD$^+$      & NeDH$^+$       & NeD$_2$$^+$   \\ 
\hline
r$_0$(Ne$-$H$_1$)  & 1.234 199 \AA & 1.231 038 \AA & 1.230 760 \AA  & 1.228 380 \AA  \\
r$_0$(H$_1-$H$_2$) & 1.105 262 \AA & 1.105 968 \AA & 1.101 848 \AA  & 1.104 113 \AA  \\
$B_0$              & 78 115.8 MHz  & 45 698.4 MHz  & 66 882.8 MHz   & 42 241.2 MHz  \\
$\alpha^B$ 1       &  2382.0 MHz   &  2412.4 MHz   &    30.2 MHz    &  931.2 MHz      \\
$\alpha^B$ 2       &  3707.2 MHz   &  1309.7 MHz   &  3274.4 MHz    &  1447.6 MHz      \\
$\alpha^B$ 3       & -1482.8 MHz   & -1054.4 MHz   &  -566.4 MHz    & -564.8 MHz      \\
$\tau_{aaaa}$      & -9.283 MHz    & -2.972 MHz    &  -7.646 MHz    & -2.665 MHz    \\
\hline                                                    
r$_e$(Ne$-$H$_1$)  & 1.211 598 \AA & --            & --             & --             \\
r$_e$(H$_1-$H$_2$) & 1.100 735 \AA & --            & --             & --             \\
$B_e$              & 79 677.6 MHz  & 46 505.0 MHz  & 67 968.6 MHz   & 42 865.8 MHz  \\
$D_e$              &  2.321 MHz    &  0.743 MHz    &  1.912 MHz     &  0.666 MHz      \\
$H_e$              & -49.088  Hz   & -3.014 Hz     & -62.171  Hz    & -7.058  Hz      \\
\hline                                                    
$\omega_1(\sigma)$ H$-$H   & 1907.4 cm$^{-1}$ & 1722.6 cm$^{-1}$ & 1634.8 cm$^{-1}$ & 1349.7 cm$^{-1}$ \\
$\omega_2(\sigma)$ Ne$-$H  &  966.4 cm$^{-1}$ &  777.6 cm$^{-1}$ &  814.1 cm$^{-1}$ &  711.4 cm$^{-1}$ \\
$\omega_3(\pi)$ Bend       &  633.5 cm$^{-1}$ &  598.7 cm$^{-1}$ &  495.2 cm$^{-1}$ &  449.9 cm$^{-1}$ \\
Harmonic Zero-Point        & 2070.4 cm$^{-1}$ & 1848.8 cm$^{-1}$ & 1719.7 cm$^{-1}$ & 1480.5 cm$^{-1}$ \\ 
$\nu_1(\sigma)$ H$-$H      & 1800.6 cm$^{-1}$ & 1610.6 cm$^{-1}$ & 1525.7 cm$^{-1}$ & 1320.2 cm$^{-1}$ \\
$\nu_2(\sigma)$ Ne$-$H     &  790.0 cm$^{-1}$ &  633.7 cm$^{-1}$ &  719.6 cm$^{-1}$ &  612.4 cm$^{-1}$ \\
$\nu_3(\pi)$ Bend          &  557.0 cm$^{-1}$ &  530.3 cm$^{-1}$ &  448.5 cm$^{-1}$ &  404.6 cm$^{-1}$ \\
Zero-Point                 & 1999.8 cm$^{-1}$ & 1791.5 cm$^{-1}$ & 1670.1 cm$^{-1}$ & 1445.5 cm$^{-1}$ \\
\hline                                                    
$B_1$       & 75 733.8 MHz  & 43 286.0 MHz  & 66 852.6 MHz   & 41 310.0 MHz  \\
$B_2$       & 74 408.5 MHz  & 44 388.6 MHz  & 63 608.4 MHz   & 40 793.6 MHz  \\
$B_3$       & 74 408.5 MHz  & 46 752.8 MHz  & 67 449.2 MHz   & 42 806.1 MHz  \\

\hline
\end{tabular}

\end{table*}
\endgroup

\renewcommand{\baselinestretch}{1}
\begingroup
\begin{table*}

\caption{The ArH$_2$$^+$ CcCR Simple-Internal Force Constants (in
mdyn/\AA$^n$$\cdot$rad$^m$)$^a$.}

\label{Arfc}

\centering

\begin{tabular}{c r c r c r c r c r}
\hline

F$_{11}$ & 1.895 279 & F$_{331}$ & -0.1878 & F$_{3311}$ & -8.29 \\ 
F$_{21}$ & 0.621 557 & F$_{332}$ & 0.0940 & F$_{3321}$ & 1.67 \\ 
F$_{22}$ & 0.700 608 & F$_{441}$ & -0.1878 & F$_{3322}$ & -46.58 \\ 
F$_{33}$ & 0.077 729 & F$_{442}$ & 0.0940 & F$_{3333}$ & -101.03 \\
F$_{44}$ & 0.077 729 & F$_{1111}$ & 56.65 & F$_{4411}$ & -8.29 \\ 
F$_{111}$ & -11.0653 & F$_{2111}$ & 1.31 & F$_{4421}$ & 1.67 \\ 
F$_{211}$ & -1.5955 &  F$_{2211}$ & -3.62 & F$_{4422}$ & -46.58 \\ 
F$_{221}$ & -0.5609 &  F$_{2221}$ & 0.64 & F$_{4433}$ & -33.57 \\ 
F$_{222}$ & -3.6640 &  F$_{2222}$ & 9.06 & F$_{4444}$ & -101.03 \\

\hline
\end{tabular}

$^a$1 mdyn $=$ $10^{-8}$ N; see Figure \ref{Nefc} footnote $a$ for more details.
\end{table*}
\endgroup
\renewcommand{\baselinestretch}{2}

\renewcommand{\baselinestretch}{1}
\begingroup
\begin{table*}

\caption{The CcCR Zero-Point ($R_{\alpha}$ vibrationally-averaged) and
Equilibrium Structures, Rotational Constants, CCSD(T)/aug-cc-pV5Z Dipole
Moment, Vibration-Rotation Interaction Constants, and Quartic (D, $\tau$) and
Sextic (H) Distortion Constants of $^{36}$ArH$_2$$^+$ and Deuterated
Isotopologues.}

\label{Ar36}

\centering

\small

\begin{tabular}{l | r r r r} 
\hline\hline

                   & ArH$_2$$^+$   & ArHD$^+$      & ArDH$^+$       & ArD$_2$$^+$   \\ 
\hline
r$_0$(Ar$-$H$_1$)  & 1.441 869 \AA & 1.442 966 \AA & 1.438 876 \AA  & 1.440121 \AA  \\
r$_0$(H$_1-$H$_2$) & 1.120 932 \AA & 1.113 211 \AA & 1.123 344 \AA  & 1.117636 \AA  \\
$B_0$              & 61 278.8 MHz  & 35 878.2 MHz  & 50 663.7 MHz   & 32 232.2 MHz  \\
$\alpha^B$ 1       &  2924.3 MHz   &  1801.9 MHz   &  1093.0 MHz    &  1060.7 MHz      \\
$\alpha^B$ 2       &  1246.7 MHz   &   839.5 MHz   &   900.6 MHz    &  497.6 MHz      \\
$\alpha^B$ 3       & -1575.0 MHz   & -1132.0 MHz   & -670.9 MHz     & -585.2 MHz      \\
$\tau_{aaaa}$      & -3.199 MHz    & -1.235 MHz    & -2.015 MHz     & -0.886 MHz    \\
\hline                                                    
r$_e$(Ar$-$H$_1$)  & 1.434 944 \AA & --            & --             & --             \\
r$_e$(H$_1-$H$_2$) & 1.110 007 \AA & --            & --             & --             \\
$B_e$              & 61 789.3 MHz  & 36 066.8 MHz  & 50 989.6 MHz   & 32 426.2 MHz   \\
$D_e$              & 0.800 MHz     & 0.309 MHz     & 0.511 MHz      & 0.221 MHz      \\
$H_e$              & 4.999 Hz      & -0.267 Hz     & 2.132  Hz      & 0.690  Hz      \\
$\mu_z$$^a$        & 4.37 D        & --            & --             & --             \\ 
\hline                                                    
$\omega_1(\sigma)$ H$-$H   & 1531.3 cm$^{-1}$ & 1529.3 cm$^{-1}$ & 1121.1 cm$^{-1}$ & 1095.7 cm$^{-1}$ \\
$\omega_2(\sigma)$ Ar$-$H  & 1096.6 cm$^{-1}$ &  787.0 cm$^{-1}$ & 1073.4 cm$^{-1}$ &  787.0 cm$^{-1}$ \\
$\omega_3(\pi)$ Bend       &  665.0 cm$^{-1}$ &  623.8 cm$^{-1}$ &  524.6 cm$^{-1}$ &  471.3 cm$^{-1}$ \\
Harmonic Zero-Point        & 1979.0 cm$^{-1}$ & 1782.0 cm$^{-1}$ & 1621.9 cm$^{-1}$ & 1412.7 cm$^{-1}$ \\
$\nu_1(\sigma)$ H$-$H      & 1229.8 cm$^{-1}$ & 1333.7 cm$^{-1}$ & 1137.6 cm$^{-1}$ &  948.5 cm$^{-1}$ \\
$\nu_2(\sigma)$ Ar$-$H     &  590.5 cm$^{-1}$ &  534.3 cm$^{-1}$ & 1044.1 cm$^{-1}$ &  591.6 cm$^{-1}$ \\
$\nu_3(\pi)$ Bend          &  655.6 cm$^{-1}$ &  613.7 cm$^{-1}$ &  508.5 cm$^{-1}$ &  435.8 cm$^{-1}$ \\
Zero-Point                 & 1905.4 cm$^{-1}$ & 1735.6 cm$^{-1}$ & 1566.3 cm$^{-1}$ & 1381.3 cm$^{-1}$ \\
\hline                                                    
$B_1$       & 58 354.6 MHz  & 34 076.2 MHz  & 49 570.7 MHz   & 31 171.4 MHz  \\
$B_2$       & 60 032.1 MHz  & 35 038.7 MHz  & 49 763.1 MHz   & 31 734.6 MHz  \\
$B_3$       & 62 853.9 MHz  & 37 010.2 MHz  & 51 334.6 MHz   & 31 756.7 MHz  \\

\hline
\end{tabular}

$^a$The ArH$_2$$^+$ coordinates (in \AA\ with the centre-of-mass at the origin)
used to generate the Born-Oppenheimer dipole moment component are: Ar, 0.000000,
0.000000, -0.095551; H$_1$, 0.000000, 0.000000, 1.339394; H$_2$, 0.000000,
0.000000, 2.449401 \\ 

\end{table*}
\endgroup

\renewcommand{\baselinestretch}{1}
\begingroup
\begin{table*}

\caption{The $^{38}$ArH$_2$$^+$ and Deuterated Isotopologues CcCR Spectroscopic
Data.}

\label{Ar38}

\centering

\small

\begin{tabular}{l | r r r r} 
\hline\hline

                   & ArH$_2$$^+$   & ArHD$^+$      & ArDH$^+$       & ArD$_2$$^+$   \\ 
\hline
r$_0$(Ar$-$H$_1$)  & 1.441 848 \AA & 1.442 947 \AA & 1.438 845 \AA  & 1.440 092 \AA  \\
r$_0$(H$_1-$H$_2$) & 1.120 940 \AA & 1.113 214 \AA & 1.123 355 \AA  & 1.117 646 \AA  \\
$B_0$              & 61 121.6 MHz  & 35 740.9 MHz  & 50 475.8 MHz   & 32 075.6 MHz  \\
$\alpha^B$ 1       &  2920.1 MHz   &  1797.3 MHz   &  1082.3 MHz    &  1058.3 MHz      \\
$\alpha^B$ 2       &  1238.6 MHz   &  832.8 MHz    &  901.1 MHz     &  491.5 MHz      \\
$\alpha^B$ 3       & -1567.2 MHz   & -1127.3 MHz   & -668.3 MHz     & -582.3 MHz      \\
$\tau_{aaaa}$      &  -3.182 MHz   & -1.225 MHz    & -2.029 MHz     & -0.877 MHz    \\
\hline                                                    
r$_e$(Ar$-$H$_1$)  & 1.434 944 \AA & --            & --             & --             \\
r$_e$(H$_1-$H$_2$) & 1.110 007 \AA & --            & --             & --             \\
$B_e$              & 61 630.1 MHz  & 35 928.6 MHz  & 50 799.2 MHz   & 32 268.2 MHz   \\
$D_e$              & 0.795 MHz     &  0.306 MHz    & 0.507 MHz      & 0.219 MHz      \\
$H_e$              & 4.972 Hz      & -0.257 Hz     & 2.094  Hz      & 0.684  Hz      \\
\hline                                                    
$\omega_1(\sigma)$ H$-$H   & 1530.4 cm$^{-1}$ & 1528.3 cm$^{-1}$ & 1121.1 cm$^{-1}$ & 1094.3 cm$^{-1}$ \\
$\omega_2(\sigma)$ Ar$-$H  & 1095.7 cm$^{-1}$ &  785.9 cm$^{-1}$ & 1071.3 cm$^{-1}$ &  785.8 cm$^{-1}$ \\
$\omega_3(\pi)$ Bend       &  664.9 cm$^{-1}$ &  623.7 cm$^{-1}$ &  524.6 cm$^{-1}$ &  471.2 cm$^{-1}$ \\
Harmonic Zero-Point        & 1978.0 cm$^{-1}$ & 1780.8 cm$^{-1}$ & 1620.8 cm$^{-1}$ & 1411.3 cm$^{-1}$ \\
$\nu_1(\sigma)$ H$-$H      & 1228.6 cm$^{-1}$ & 1332.4 cm$^{-1}$ & 1139.2 cm$^{-1}$ &  946.8 cm$^{-1}$ \\
$\nu_2(\sigma)$ Ar$-$H     &  591.6 cm$^{-1}$ &  534.0 cm$^{-1}$ & 1042.0 cm$^{-1}$ &  591.3 cm$^{-1}$ \\
$\nu_3(\pi)$ Bend          &  655.6 cm$^{-1}$ &  613.7 cm$^{-1}$ &  508.7 cm$^{-1}$ &  435.8 cm$^{-1}$ \\
Zero-Point                 & 1904.5 cm$^{-1}$ & 1734.5 cm$^{-1}$ & 1565.1 cm$^{-1}$ & 1380.0 cm$^{-1}$ \\
\hline                                                    
$B_1$       & 58 201.5 MHz  & 33 943.6 MHz  & 49 393.5 MHz   & 31 017.3 MHz  \\
$B_2$       & 59 883.0 MHz  & 34 908.1 MHz  & 49 574.7 MHz   & 31 584.1 MHz  \\
$B_3$       & 62 692.5 MHz  & 36 868.3 MHz  & 51 144.1 MHz   & 32 657.9 MHz  \\

\hline
\end{tabular}

\end{table*}
\endgroup

\renewcommand{\baselinestretch}{1}
\begingroup
\begin{table*}

\caption{The CcCR Spectroscopic Data for $^{40}$ArH$_2$$^+$ and Its Deuterated
Isotopologues.}

\label{Ar40}

\centering

\small

\begin{tabular}{l | r r r r} 
\hline\hline

                   & ArH$_2$$^+$   & ArHD$^+$      & ArDH$^+$       & ArD$_2$$^+$   \\ 
\hline
r$_0$(Ar$-$H$_1$)  & 1.441 829 \AA & 1.442 930 \AA & 1.438 816 \AA  & 1.440 066 \AA  \\
r$_0$(H$_1-$H$_2$) & 1.120 946 \AA & 1.113 216 \AA & 1.123 364 \AA  & 1.117 656 \AA  \\
$B_0$              & 60 979.5 MHz  & 35 616.8 MHz  & 50 306.0 MHz   & 31 934.2 MHz  \\
$\alpha^B$ 1       &  2916.4 MHz   &  1793.1 MHz   &  1073.1 MHz    &  1056.0 MHz      \\
$\alpha^B$ 2       &  1231.4 MHz   &  826.8 MHz    &  901.3 MHz     &  486.1 MHz      \\
$\alpha^B$ 3       & -1567.2 MHz   & -1123.1 MHz   & -666.0 MHz     & -579.7 MHz      \\
$\tau_{aaaa}$      & -3.166 MHz    & -1.216 MHz    & -2.015 MHz     & -0.869 MHz    \\
\hline                                                    
r$_e$(Ar$-$H$_1$)  & 1.434 944 \AA & --            & --             & --             \\
r$_e$(H$_1-$H$_2$) & 1.110 007 \AA & --            & --             & --             \\
$B_e$              & 61 486.2 MHz  & 35 803.6 MHz  & 50 627.2 MHz   & 32 125.5 MHz   \\
$D_e$              & 0.792 MHz     & 0.304 MHz     & 0.504 MHz      & 0.217 MHz      \\
$H_e$              & 4.947 Hz      & -0.248 Hz     & 2.061  Hz      & 0.678  Hz      \\
\hline                                                    
$\omega_1(\sigma)$ H$-$H   & 1529.6 cm$^{-1}$ & 1527.4 cm$^{-1}$ & 1121.1 cm$^{-1}$ & 1093.1 cm$^{-1}$ \\
$\omega_2(\sigma)$ Ar$-$H  & 1094.9 cm$^{-1}$ &  784.9 cm$^{-1}$ & 1069.3 cm$^{-1}$ &  784.8 cm$^{-1}$ \\
$\omega_3(\pi)$ Bend       &  664.9 cm$^{-1}$ &  623.7 cm$^{-1}$ &  524.5 cm$^{-1}$ &  471.2 cm$^{-1}$ \\
Harmonic Zero-Point        & 1977.2 cm$^{-1}$ & 1779.9 cm$^{-1}$ & 1619.7 cm$^{-1}$ & 1410.2 cm$^{-1}$ \\
$\nu_1(\sigma)$ H$-$H      & 1214.8 cm$^{-1}$ & 1331.2 cm$^{-1}$ & 1140.5 cm$^{-1}$ &  945.2 cm$^{-1}$ \\
$\nu_2(\sigma)$ Ar$-$H     &  592.5 cm$^{-1}$ &  533.7 cm$^{-1}$ & 1040.2 cm$^{-1}$ &  591.1 cm$^{-1}$ \\
$\nu_3(\pi)$ Bend          &  655.6 cm$^{-1}$ &  613.6 cm$^{-1}$ &  508.8 cm$^{-1}$ &  435.8 cm$^{-1}$ \\
Zero-Point                 & 1904.0 cm$^{-1}$ & 1733.6 cm$^{-1}$ & 1564.0 cm$^{-1}$ & 1378.8 cm$^{-1}$ \\
\hline                                                    
$B_1$       & 58 063.1 MHz  & 33 823.7 MHz  & 49 232.9 MHz   & 30 878.2 MHz  \\
$B_2$       & 59 748.1 MHz  & 34 790.0 MHz  & 51 849.6 MHz   & 31 448.1 MHz  \\
$B_3$       & 62 546.7 MHz  & 36 739.8 MHz  & 50 972.0 MHz   & 32 513.9 MHz  \\

\hline
\end{tabular}

\end{table*}
\endgroup

\renewcommand{\baselinestretch}{1}
\begingroup
\begin{table*}

\caption{CCSD(T)/aug-cc-pVTZ Dissociation Pathways of NeH$_2$$^+$ and
ArH$_2$$^+$ with Energies Relative to the Respective Triatomic Cation Minima.} 

\label{Diss}

\centering

\begin{tabular}{l l | l l}
\hline

NeH$_2$$^+$ Products & Relative Energy & ArH$_2$$^+$ Products & Relative Energy\\
\hline
NeH$^+$ + H    & 1.031 eV & ArH$^+$ + H    & 0.493 eV \\
Ne + H$_2$$^+$ & 0.560 eV & Ar + H$_2$$^+$ & 1.838 eV \\
Ne$^+$ + H$_2$ & 6.503 eV & Ar$^+$ + H$_2$ & 1.930 eV \\
Ne + H + H$^+$ & 3.349 eV & Ar + H + H$^+$ & 4.627 eV \\
 
\hline
\end{tabular}

\end{table*}
\endgroup
\renewcommand{\baselinestretch}{2}

\renewcommand{\baselinestretch}{1}
\begingroup
\begin{table*}
\caption{The CcCR Pure Rotational Limited Line List for ArH$_2$$^+$ with
$\Delta E_{J+1 \rightarrow J}, J \leq 9$ with Frequencies in MHz and Intensities
in MHz (per molecule).},
\label{LLL}
\centering
\small
\begin{tabular}{l c c c c c}
\hline\hline

Molecule & J+1 & J & Frequency & 5000K Intensity & 40K Intensity \\
$^{36}$ArH$_2$$^+$ & 1 & 0 & 122554.2 & 8.010$\times 10^{-08}$ & 0.001194 \\
$^{36}$ArH$_2$$^+$ & 2 & 1 & 245087.3 & 3.200$\times 10^{-07}$ & 0.003841 \\
$^{36}$ArH$_2$$^+$ & 3 & 2 & 367584.5 & 7.170$\times 10^{-07}$ & 0.006010 \\
$^{36}$ArH$_2$$^+$ & 4 & 3 & 490021.7 & 1.270$\times 10^{-06}$ & 0.006426 \\
$^{36}$ArH$_2$$^+$ & 5 & 4 & 612384.0 & 1.970$\times 10^{-06}$ & 0.005223 \\
$^{36}$ArH$_2$$^+$ & 6 & 5 & 734650.4 & 2.820$\times 10^{-06}$ & 0.003384 \\
$^{36}$ArH$_2$$^+$ & 7 & 6 & 856799.8 & 3.810$\times 10^{-06}$ & 0.001793 \\
$^{36}$ArH$_2$$^+$ & 8 & 7 & 978817.3 & 4.930$\times 10^{-06}$ & 0.000789 \\
$^{36}$ArH$_2$$^+$ & 9 & 8 & 1100682.0 & 6.180$\times 10^{-06}$ & 0.000291 \\
$^{36}$ArH$_2$$^+$ & 10 & 9 & 1222372.7 & 7.540$\times 10^{-06}$ & 0.000091 \\
\hline
$^{38}$ArH$_2$$^+$ & 1 & 0 & 122239.4 & 2.905$\times 10^{-09}$ & 0.000043 \\
$^{38}$ArH$_2$$^+$ & 2 & 1 & 244457.8 & 1.160$\times 10^{-08}$ & 0.000139 \\
$^{38}$ArH$_2$$^+$ & 3 & 2 & 366640.2 & 2.602$\times 10^{-08}$ & 0.000218 \\
$^{38}$ArH$_2$$^+$ & 4 & 3 & 488765.6 & 4.606$\times 10^{-08}$ & 0.000234 \\
$^{38}$ArH$_2$$^+$ & 5 & 4 & 610813.1 & 7.157$\times 10^{-08}$ & 0.000190 \\
$^{38}$ArH$_2$$^+$ & 6 & 5 & 732767.7 & 1.024$\times 10^{-07}$ & 0.000124 \\
$^{38}$ArH$_2$$^+$ & 7 & 6 & 854605.4 & 1.382$\times 10^{-07}$ & 0.000066 \\
$^{38}$ArH$_2$$^+$ & 8 & 7 & 976311.1 & 1.789$\times 10^{-07}$ & 0.000029 \\
$^{38}$ArH$_2$$^+$ & 9 & 8 & 1097864.0 & 2.241$\times 10^{-07}$ & 0.000011 \\
$^{38}$ArH$_2$$^+$ & 10 & 9 & 1219245.9 & 2.735$\times 10^{-07}$ & 0.000003 \\

\hline
\end{tabular}

\end{table*}
\endgroup

\end{document}